\documentclass[11pt,a4paper,floats]{article}

\usepackage{graphics}
\usepackage{epsf}
\usepackage{epsfig}
\begin{document} 

\flushbottom

\title{\flushleft Universal scaling behavior of directed percolation\\
around the upper critical dimension}




\date{}

\maketitle






\setcounter{page}{1}
\markright{Journal of Statistical Physics {\textbf 115}, 1231 (2004)}
\thispagestyle{myheadings}
\pagestyle{myheadings}


\noindent{\textbf {S.~L\"ubeck\footnote{
Weizmann Institute of Science, 
Department of Physics of Complex Systems, 
76100 Rehovot, Israel,
}
\footnote{Institut f\"ur Theoretische Physik,
Universit\"at Duisburg-Essen, 
47048 Duisburg, Germany}
and R.\,D.~Willmann\footnote{
Institut f\"ur Festk\"orperforschung,
Forschungszentrum J\"ulich, 52425 J\"ulich, Germany
}}}\\[3mm]

\hfill\parbox{9.7cm}{
\footnotesize
{\textit {Received October 20, 2003; accepted Dezember 28, 2003}}\\[-2mm]
\hrule
\vspace{1mm}
In this work we consider the steady state scaling behavior of 
directed percolation around the upper critical dimension.
In particular we determine numerically 
the order parameter, its fluctuations as
well as the susceptibility as a function of the control
parameter and the conjugated field.
Additionally to the universal scaling functions,  
several universal amplitude combinations are considered.
We compare our results with those of a renormalization group
approach.\\[-2mm]
\hrule
\vspace{2mm}
}

\vspace{10mm}

\section{\textsf {INTRODUCTION}}

The concept of universality is one of the most impressive 
features of continuous phase transitions.
It allows to group the great variety of models into a small 
number of universality 
classes~(see~\cite{STANLEY_1} for a recent review). 
Models within one class share the same critical exponents. 
Furthermore their corresponding scaling functions become identical 
close to the critical point. 
Often, universality classes are also characterized by certain
amplitude combinations, which are merely particular
values of the scaling functions.
The most prominent examples of universal behavior are 
the coexistence curve of liquid-vapor systems~\cite{GUGGENHEIM_1} 
and the equation of state in ferromagnetic 
systems~(e.g.~\cite{STANLEY_1,MILOSEVIC_2}).
Deciding on a systems universality class by
considering the scaling functions instead of critical
exponents appears to be less prone to errors in most cases.
While for the latter ones the variations between 
different universal classes are often small, the amplitude
combinations and therefore the scaling functions may differ
significantly (see~\cite{PRIVMAN_2}).


Wilsons renormalization group (RG) approach lays the 
foundation for an understanding of universality~\cite{WILSON_1,WILSON_2}. 
It also yields a tool for computing critical exponents 
as well as the universal scaling functions. 
While critical exponents emerge from local properties
near a given fixed point, scaling functions require
the knowledge of the full RG flow along the trajectories between 
neighboring fixed points.
This illustrates why scaling functions are more
{\it sensitive} than the corresponding exponents.

The RG explains the existence of an 
upper critical dimension~$D_{\scriptscriptstyle \mathrm c}$ above which the 
mean-field theory applies, i.e., classical theories,
which neglect strong fluctuations and correlations, 
provide correct estimates to the critical exponents and scaling
functions.
Below $D_{\scriptscriptstyle  \mathrm c}$ fluctuations 
become relevant and
the mean-field scenario breaks down.
At the upper critical dimension the RG equations yield
mean-field exponents with logarithmic corrections~\cite{WEGNER_1,WILSON_3}.


In contrast to equilibrium critical phenomena less is known
in the case of non-equilibrium phase transitions.
This is due to the fact that a generalized
treatment is not possible, lacking an analog to the
equilibrium free energy.
The rich and often surprising variety of phenomena
has to be studied for each system individually.
The scaling behavior of directed percolation (DP) is recognized 
as the paradigm of the critical behavior
of several non-equilibrium systems which exhibit a 
continuous phase transition from an active 
to an absorbing state (see e.g.~\cite{HINRICHSEN_1}).
The widespread occurrence of such models describing 
phenomena in physics, biology, as well as 
catalytic chemical reactions is reflected by the 
well known universality hypothesis of Janssen and Grassberger:
Short-range interacting models, which 
exhibit a continuous phase transition into a unique 
absorbing state belong to the DP universality class,
provided they have a one-component order parameter
and no additional symmetries~\cite{JANSSEN_1,GRASSBERGER_2}.
Different universality classes are expected to occur 
in the presence of additional symmetries,
like particle conservation~\cite{ROSSI_1}, 
particle-hole symmetry (compact directed percolation)~\cite{ESSAM_1}, or
parity conservation (e.g.~branching annihilating random walks with an
even number of offsprings~\cite{CARDY_2}).
Other model details, such as e.g.~the geometry or shape of a lattice,
are expected to have no
influence on the scaling behavior in the vicinity of the 
critical point.

The universality hypothesis still awaits a rigorous proof.
Amazingly, numerous simulations suggest that the DP 
universality class is even larger than expected.
It turns out that the hypothesis defines only a sufficient
condition but fails to describe the DP class
in full generality (see~\cite{HINRICHSEN_1} for a detailed 
discussion).
For instance, the pair contact process (PCP)
is one of the simplest models with infinitely many
absorbing states exhibiting a continuous phase 
transition~\cite{JENSEN_2}.
It was shown that the critical scaling behavior of the
one-dimensional PCP is characterized by the same critical 
exponents~\cite{JENSEN_2,JENSEN_3} as well
as by the same universal scaling functions as DP~\cite{LUEB_27}.
Thus despite the different structure of the absorbing phase
the one-dimensional PCP belongs to the DP universality class.
This numerical evidence confirms a corresponding 
RG-conjecture~\cite{MUNOZ_1}.
But one has to mention that a recently performed
RG analysis conjectures a different scaling behavior
of both models in higher dimensions~\cite{WIJLAND_1}.


In this work we consider the universal scaling behavior of
directed percolation in various dimensions.
Whereas most investigations on DP follow the seminal 
work ref.~\cite{GRASSBERGER_4} 
and thus focus on activity spreading we examine
the steady state scaling behavior for $D\ge 2$.
We determine the universal scaling functions
of the order parameter (i.e.~the equation of state) and its
fluctuations.
Furthermore we consider certain universal amplitude combinations
which are related to the order parameter and its
susceptibility.
These amplitude combinations are immediately related 
to particular values of the universal scaling functions
and are of great experimental interest~\cite{PRIVMAN_2}.
We will see that the numerically obtained universal 
scaling functions and the related universal amplitude combinations
allow a quantitative test of RG-results.
The powerful and versatile $\epsilon$-expansion 
provide estimates of almost all quantities of interest, 
e.g.~the critical exponents and the scaling functions 
(see e.g.~\cite{PFEUTY_1}).
Unfortunately it is impossible to 
estimate within this approximation scheme
the corresponding error-bars.
Thus it is intriguing to compare our results with those of 
RG analysis~\cite{JANSSEN_2,JANSSEN_3}.

Furthermore we focus on the phase transition at the
upper critical dimension $D_{\scriptscriptstyle \mathrm c}=4$.
There the usual power-laws are modified by logarithmic 
corrections.
These logarithmic corrections are well 
established in equilibrium critical phenomena~\cite{WEGNER_1,WILSON_3}
but they have been largely ignored for non-equilibrium
phase transitions.
Due to the considerable numerical effort,
sufficiently accurate simulation data for non-equilibrium
systems became available only recently:
Investigated systems include
self-avoiding random walks~\cite{GRASSBERGER_6,GRASSBERGER_7},
self-organized critical systems~\cite{LUEB_5,LUEB_10},
depinning-transitions in disordered media~\cite{LUEB_17}, 
isotropic percolation~\cite{GRASSBERGER_5},
as well as absorbing phase transitions~\cite{LUEB_26}.
On the other hand, the numerical advance triggered further
analytical RG calculations yielding estimates for 
the logarithmic correction exponents for the respective 
systems~\cite{FEDORENKO_1,JANSSEN_5,STENULL_1,JANSSEN_3}.


The outline of the present paper is as follows:~The next
section contains the model definition and a description
of the method of numerical analysis.
In Sec.\,\ref{sec:uni_scal_forms} we describe the 
scaling behavior at the critical point and introduce
the critical exponents as well as the 
universal scaling functions.
The numerical data of the order parameter and its
fluctuations are analyzed in Sec.\,\ref{sec:eqos_fluc}
below ($D=2,3$), above ($D=5$), and at the upper critical
dimension ($D=4$).
Several amplitude combinations are considered in
Sec.\,\ref{sec:uni_ampl_comb}.
Concluding remarks are given in Sec.\,\ref{sec:conc}.

\section{\textsf {MODEL AND SIMULATIONS}}
\label{sec:mod_sim}

In order to examine the scaling behavior of the $D$-dimensional 
DP universality class we consider the directed site percolation
process using a generalized
Domany-Kinzel automaton~\cite{DOMANY_1}.
It is defined on a $D+1$-dimensional body centered cubic (bcc) 
lattice (where time corresponds to the $[0,0,\ldots,0,1]$
direction) and evolves 
by parallel update according to the following 
rules:~A site at time~$t$ is occupied with 
probability~$p$ if at least one of its $2^{D}$ backward
neighboring sites (time $t-1$) is occupied.
Otherwise the site remains empty.
Furthermore, spontaneous particle creation may take
place at all sites with probability $p_{\scriptscriptstyle 0}$.
Directed site percolation corresponds to the choice
$p_{\scriptscriptstyle 0}=0$.
The propagation probability~$p$ is the control parameter
of the phase transition, i.e., below a 
critical value~$p_{\scriptscriptstyle\mathrm c}$ the activity ceases 
and the system is trapped forever in the absorbing state
(empty lattice).
On the other hand a non-zero density of (active) 
particles $\rho_{\scriptscriptstyle\mathrm a}$ is
found for $p>p_{\scriptscriptstyle\mathrm c}$.
The best estimates of the critical value of directed site
percolation on bcc lattices are 
$p_{\scriptscriptstyle \mathrm c}=0.705489(4)$~\cite{TRETYAKOV_1} for $D=1$
and 
$p_{\scriptscriptstyle \mathrm c}=0.34457(1)$~\cite{GRASSBERGER_3} for $D=2$.

\begin{figure}[b]
  \centering
  \includegraphics[width=7cm,angle=0]{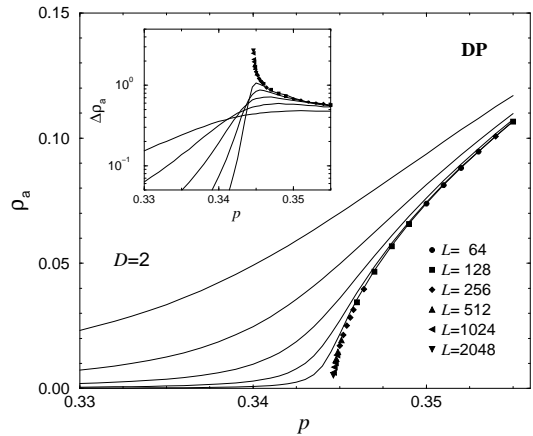}
  \caption{
    The two-dimensional directed percolation 
    order parameter~$\rho_{\scriptscriptstyle \rm a}$ as a function of the
    particle density for zero field (symbols) and for various 
    values of the external field 
    ($h=3\,10^{-4}$, $10^{-4}$, $2\,10^{-5}$, $5\,10^{-6}$, $10^{-6}$) (lines).
    The inset displays the order parameter 
    fluctuations~$\Delta \rho_{\scriptscriptstyle \rm a}$ 
    for zero field (symbols)
    and for various values of the external field $h$ (lines). 
   }
  \label{fig:rho_a_2d_01} 
\end{figure}

The order parameter $\rho_{\scriptscriptstyle \mathrm{a}}$ of the 
absorbing phase transition  
vanishes at the critical point according to
\begin{equation}
\label{eq:def_beta}
\rho_{\scriptscriptstyle \mathrm{a}} \; \propto \; \delta p^{\beta},
\end{equation}
with $\delta p=(p-p_{\scriptscriptstyle\mathrm c})/p_{\scriptscriptstyle \mathrm c}$.
Furthermore the order parameter fluctuations 
$\Delta\rho_{\scriptscriptstyle \mathrm{a}} 
= L^{D} ( \langle \rho_{\scriptscriptstyle \mathrm{a}}^2 \rangle - \langle
\rho_{\scriptscriptstyle \mathrm{a}}\rangle^2)$  diverge as
\begin{equation}
\label{eq:def_sigma_prime}
\Delta \rho_{\scriptscriptstyle \mathrm{a}} \; \propto \; \delta p^{- \gamma^{\prime}}.
\end{equation} 
The fluctuation exponent $\gamma^{\prime}$ obeys the scaling relation 
$\gamma^{\prime}=D \nu_{\scriptscriptstyle \perp}-2 \beta$~\cite{JENSEN_3}, 
where $\nu_{\scriptscriptstyle \perp}$ 
describes the divergence of 
the spatial correlation length at the critical point.
The critical behavior of the order parameter is shown in
Fig.\,\ref{fig:rho_a_2d_01} for $D=2$.
The data are obtained from numerical simulations of 
systems with periodic boundary conditions.
Considering various system sizes~$L$ we 
take care that our results are not affected by finite-size
effects.
The system is started from a random initial configuration.
After a certain transient regime a steady state is reached,
which is characterized by the average particle 
density $\rho_{\scriptscriptstyle \mathrm a}$
and its fluctuations~$\Delta\rho_{\scriptscriptstyle \mathrm a}$.

Similar to equilibrium phase transitions it is possible
in DP to apply an external field~$h$ that is conjugated
to the order parameter.
Being a conjugated field its has to destroy
the absorbing phase and the corresponding
linear response function has to diverge at the critical
point i.e.,
\begin{equation}
\label{eq:suscept_behavior}
\chi_{\scriptscriptstyle \mathrm a}  
\; =  \; \frac{\partial\rho_{\mathrm a}}{\partial h} 
\; \to \; \infty.
\end{equation}
In DP the external field is 
implemented~\cite{HINRICHSEN_1,LUEB_27} as a spontaneous
creation of particles (i.e.~$p_{\scriptscriptstyle 0}=h>0$).
Clearly, the absorbing state and thus the phase transition
are destroyed. 
Figure\,\ref{fig:rho_a_2d_01} shows how the external field
results in a smoothening of the zero-field order parameter curve.
The inset displays that the fluctuations are peaked
for finite fields. 
Approaching the transition point ($h\to 0$) this peak
becomes a divergence signalling the critical point.

\section{\textsf {UNIVERSAL SCALING FORMS}}
\label{sec:uni_scal_forms}

Sufficiently close to the critical point
the order parameter,
its fluctuations, as well as the order parameter susceptibility
can be described by generalized homogeneous functions
\begin{eqnarray}
\label{eq:scal_ansatz_EqoS}
\rho_{\scriptscriptstyle \mathrm a}(\delta p, h) 
\; & \sim & \; 
\lambda^{-\beta}\, \, {\tilde R}
(a_{\scriptscriptstyle p}  
\delta p \; \lambda, a_{\scriptscriptstyle h} h \;
\lambda^{\sigma}) \, ,\\[2mm]
\label{eq:scal_ansatz_Fluc}
a_{\scriptscriptstyle \Delta} \,
\Delta \rho_{\scriptscriptstyle \mathrm a}(\delta p, h) 
\; & \sim & \; 
\lambda^{\gamma^{\prime}}\, \, {\tilde D}
(a_{\scriptscriptstyle p} \delta p \; \lambda, 
a_{\scriptscriptstyle h} h \, \lambda^{\sigma})  \, , \\[2mm]
\label{eq:scal_ansatz_Susc}
a_{\scriptscriptstyle \chi} \,
\chi_{\scriptscriptstyle \mathrm a}(\delta p, h) 
\; & \sim & \; 
\lambda^{\gamma}\, \, {\tilde C}
(a_{\scriptscriptstyle p} \delta p \; \lambda, 
a_{\scriptscriptstyle h} h \, \lambda^{\sigma})  \, .
\end{eqnarray}
Note that these scaling forms are valid 
for $D\neq D_{\scriptscriptstyle\mathrm c}$.
At the upper critical dimension $D_{\scriptscriptstyle\mathrm c}$
they have to be modified by logarithmic corrections~\cite{LUEB_26}.
Taking into consideration that the susceptibility is defined as the
derivative of the order parameter with respect to the 
conjugated field [Eq.\,(\ref{eq:suscept_behavior})]
we find ${\tilde C}(x,y)=\partial_y {\tilde R}(x,y)$,
$a_{\scriptscriptstyle \chi}= a_{\scriptscriptstyle h}^{-1}$,
as well as the Widom scaling law
\begin{equation}
\gamma \; = \; \sigma  \, - \, \beta \, .
\label{eq:widom}
\end{equation}

The universal scaling functions ${\tilde R}$, ${\tilde D}$,
and ${\tilde C}$
are identical for all models belonging to a given
universality class whereas all non-universal
system-dependent details (e.g.~the lattice structure,
range of interactions, the update scheme, etc.) are 
contained in the so-called
non-universal metric factors
$a_{\scriptscriptstyle p}$, 
$a_{\scriptscriptstyle h}$, 
and $a_{\scriptscriptstyle \Delta}$~\cite{PRIVMAN_3}.
The universal scaling functions can be normalized
by the conditions ${\tilde R}(1,0)={\tilde R}(0,1)={\tilde D}(0,1)=1$.
In that case the non-universal metric factors are 
determined by the amplitudes of the corresponding
power-laws
\begin{eqnarray}
\label{eq:metric_factors_a_rho}
\rho_{\scriptscriptstyle \mathrm a}(\delta p, h=0) \; & \sim & \; 
(a_{\scriptscriptstyle p} \, \delta p)^{\beta} \, ,\\
\label{eq:metric_factors_a_h}   
\rho_{\scriptscriptstyle \mathrm a}(\delta p =0, h) \; & \sim & \; 
(a_{\scriptscriptstyle h} \, h)^{\beta / \sigma} \, , \\
\label{eq:metric_factors_a_Delta}   
a_{\scriptscriptstyle \Delta} \,
\Delta\rho_{\scriptscriptstyle \mathrm a}(\delta p=0, h) \; & \sim &\; 
(a_{\scriptscriptstyle h} \, h)^{-\gamma^{\prime}/\sigma} \, .
\end{eqnarray}
Furthermore we just mention that  
${\tilde C}(0,1)= \beta / \sigma$, 
following trivially from the definition of the
susceptibility.

\begin{table}[b]
\caption{The non-universal quantities for site directed percolation
on a bcc lattice for various dimensions.
The uncertainty of the metric factors is less than 7\%.
The values for $D=1$ are obtained from a previous work~\protect\cite{LUEB_27}.}
\label{table:metric_factors}
\centering
\begin{tabular}{llllll}
$D\;\;$ & $p_{\scriptscriptstyle \mathrm c}$ &
$a_{\scriptscriptstyle \mathrm a}$  & 
$a_{\scriptscriptstyle p}$     &
$a_{\scriptscriptstyle h}$        &
$a_{\scriptscriptstyle \Delta}$   \\
\hline \\
$1$     & $0.705489\pm0.000004\;\;$ & $$                & $2.498$       & $0.114$       &
$9.382$ \\
$2$     & $0.344575\pm0.000015\;\;$ & $$                & $0.795$       & $0.186$       &
$9.016$ \\
$3$     & $0.160950\pm0.000030$     & $$                & $0.417$       & $0.328$       &
$11.91$ 
\\
$4$     & $0.075582\pm0.000017$     & $14.70\;$ & $3.055$       & $59.80$       & $19.19$ 
\\
$5$     & $0.035967\pm0.000023$     & $$                & $0.114\;$     & $0.174\;$     &
$42.49\;$
\end{tabular}
\end{table}

Usually, an analytical expression for the scaling
functions is only known above $D_{\scriptscriptstyle \mathrm c}$,
where mean-field theories apply.
In the case of directed percolation the mean-field
scaling functions are given 
by~(see e.g.~\cite{MORI_1})
\begin{eqnarray}
\label{eq:uni_scal_mf_R}
{\tilde R}_{\scriptscriptstyle \mathrm {MF}}  
(x, y) & = &
\frac{x}{2} \, + \, \sqrt{y \, + \,
\left (\frac{x}{2} \right )^2 \;} \, ,   \\
\label{eq:uni_scal_mf_D}
{\tilde D}_{\scriptscriptstyle \mathrm {MF}}  
(x , y) & = &
\frac{{\tilde R}_{\scriptscriptstyle \mathrm {MF}}(x,y)}
{\,{\sqrt{y \, + \,\left ( {x}/{2} \right )^2 \;}}\,} \, , \\
\label{eq:uni_scal_mf_C}
{\tilde C}_{\scriptscriptstyle \mathrm {MF}}  
(x , y) & = & 
\frac{1}{\,2\, \,{\sqrt{y \, + \,\left ( {x}/{2} \right )^2 \;}}\,} \, , 
\end{eqnarray}
i.e., the mean-field exponents are 
$\beta_{\scriptscriptstyle \mathrm {MF}} =1$, 
$\sigma_{\scriptscriptstyle \mathrm {MF}}=2$, 
$\gamma_{\scriptscriptstyle \mathrm {MF}}=1 $, and
$\gamma^{\prime}_{\scriptscriptstyle \mathrm {MF}} =0$
(corresponding to a finite jump of the fluctuations).
Below $D_{\scriptscriptstyle \mathrm c}$ the universal
scaling functions depend on dimensionality and 
are unknown due to a lack of analytical solutions.
In this case the scaling functions have to be determined 
numerically or via approximation schemes, e.g.~series
expansions or $\epsilon$-expansion of RG approaches.

\begin{table}[t]
\centering
\caption{The critical exponents of directed percolation
for various dimensions~$D$.
The one-dimensional values were obtained in a famous
series expansion by Jensen~\protect\cite{JENSEN_5}.
For $D=2$ and $D=3$ the authors investigated
activity spreading and the presented
exponents are derived via scaling relations.
A complete list of all critical exponents of DP
can be found in~\protect\cite{HINRICHSEN_1}.
The symbol $^{\ast}$ denotes logarithmic corrections
to the power-law behavior.}
\label{table:critical_indicees}
\begin{tabular}{llllll}
$\;D$       
& $\;1\;$\protect\cite{JENSEN_5} 
& $\;2\;$\protect\cite{VOIGT_1}        
& $\;3\;$\protect\cite{JENSEN_6}        
& $4$ & MF \\  
\hline \\
$\;\beta\;$ &  $\;0.276486(8)\;$ & $\;0.584(4)\;$ & $\;0.81(1)\;$  &
$\;1^{\ast}\;$ & $\;1$ \\  
$\;\sigma\;$ &  $\;2.554216(13)\;$ & $\;2.18(1)\;$ & $\;2.04(2)\;$  &
$\;2^{\ast}\;$ & $\;2$ \\  
$\;\gamma^{\prime}\;$ &  $\;0.543882(16)\;$ & $\;0.300(11)\;$ & $\;0.123(25)\;$  &
$\;0^{\ast}\;$ & $\;0$ \\  
\end{tabular}
\end{table}

In case of the mean-field 
solution ($\gamma^{\prime}_{\scriptscriptstyle {\mathrm {MF}}}=0$)
the scaling form of the fluctuations [Eq.\,(\ref{eq:scal_ansatz_Fluc})]
reduces to 
\begin{equation}
a_{\scriptscriptstyle \Delta} \, 
\Delta \rho_{\mathrm a}(\delta p, h) 
\; \sim \; {\tilde D}
(a_{\scriptscriptstyle p} \delta p \; \lambda, 
a_{\scriptscriptstyle h} h \, \lambda^{\sigma}) .
\label{eq:uni_fluc_scal_fluc_mf}
\end{equation}
Some interesting properties of the universal 
scaling function ${\tilde D}$ can be derived 
from this form.
The non-universal metric factor 
$a_{\scriptscriptstyle \Delta}$ is determined
by 
\begin{equation}
a_{\scriptscriptstyle \Delta} \; = \;
\frac{1}{\, \Delta \rho_{\mathrm a}(\delta p=0, h)\,} 
\label{eq:mf_fluc_d}
\end{equation}
using that ${\tilde D}(0,1)=1$.
The value ${\tilde D}(1,0)$ is then given
by 
\begin{equation}
{\tilde D}(1,0) \; = \;
\frac{\, \Delta \rho_{\scriptscriptstyle \mathrm a}(\delta p, h=0) \, }
{\, \Delta \rho_{\scriptscriptstyle \mathrm a}(\delta p=0, h) \, } \, .
\label{eq:mf_fluc_D}
\end{equation}
Finally, it is worth mentioning that the mean-field scaling
function ${\tilde D}$ fulfills the symmetries
\begin{equation}
{\tilde D}(1,x)={\tilde D}(x^{-1/\sigma},1)
\label{eq:symm_1}
\end{equation}
\begin{equation}
{\tilde D}(x,1)={\tilde D}(1,x^{-\sigma})
\label{eq:symm_2}
\end{equation}
for all positive $x$.
In particular we obtain for $x \to 0$ 
${\tilde D}(1,0)={\tilde D}(\infty,1)$
and ${\tilde D}(0,1)={\tilde D}(1,\infty)$,
respectively.

\section{\textsf {EQUATION OF STATE AND FLUCTUATIONS}}
\label{sec:eqos_fluc}

\subsection{\textsf {Below the upper critical dimension}}

The scaling forms Eqs.\,(\ref{eq:scal_ansatz_EqoS}-\ref{eq:scal_ansatz_Susc})
imply that curves corresponding to different values
of the conjugated field collapse to 
the universal functions ${\tilde R}(x,1)$,
${\tilde D}(x,1)$, ${\tilde C}(x,1)$, if  
$\rho_{\scriptscriptstyle \mathrm a}\,  (a_{\scriptscriptstyle h} h)^{-\beta/\sigma}$,
$a_{\scriptscriptstyle \Delta} \Delta\rho_{\scriptscriptstyle \mathrm a}\,  
(a_{\scriptscriptstyle h} h)^{\gamma^{\prime}/\sigma}$,
and $a_{\scriptscriptstyle \chi} \chi_{\scriptscriptstyle \mathrm a}\,  
(a_{\scriptscriptstyle h} h)^{\gamma/\sigma}$
are considered as functions of the rescaled control parameter
$a_{\scriptscriptstyle p}\delta p \, (a_{\scriptscriptstyle h} h)^{-1/\sigma}$.
In a first step, the non-universal metric factors 
$a_{\scriptscriptstyle p}$, $a_{\scriptscriptstyle h}$,
$a_{\scriptscriptstyle \Delta}$ are obtained from measuring
the power-laws
Eqs.\,(\ref{eq:metric_factors_a_rho}-\ref{eq:metric_factors_a_Delta})
(see Table\,\ref{table:metric_factors}).
Here, the best known estimates for critical exponents, as given
in Table\,\ref{table:critical_indicees}, are used.


\begin{figure}[t]
  \centering
  \includegraphics[width=7cm,angle=0]{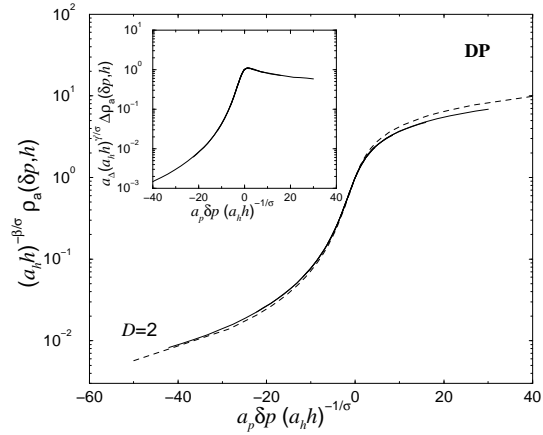}
  \caption{
    The universal scaling plots of the order parameter 
    and its fluctuations (inset) for $D=2$. 
    The dashed line corresponds to an $\epsilon$-expansion 
    of a RG approach~\protect\cite{JANSSEN_2}.
   }
  \label{fig:uni_scal_2d} 
\end{figure}

\begin{figure}[t]
  \centering
  \includegraphics[width=7cm,angle=0]{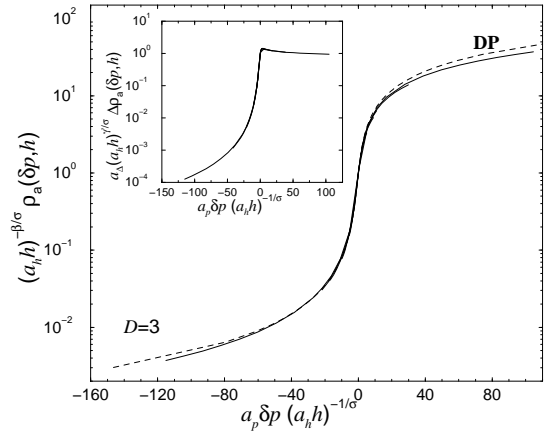}
  \caption{
    The universal scaling plots of the order parameter 
    and its fluctuations (inset) for $D=3$ and
    for various values of the external field 
    ($h=3\,10^{-4}$, $10^{-4}$, $2\,10^{-5}$, $4\,10^{-6}$, $5\,10^{-7}$).
    The dashed line corresponds to an $\epsilon$-expansion 
    of a RG approach~\protect\cite{JANSSEN_2}.
   }
  \label{fig:uni_scal_3d} 
\end{figure}

Subsequently, the rescaled order parameter and its fluctuations
as a function of the rescaled control parameter are 
plotted for two- and three-dimensional DP 
(Figs.\,\ref{fig:uni_scal_2d},\ref{fig:uni_scal_3d}).
A convincing data collapse is achieved, 
confirming the scaling ansatz as well as
the values of the critical exponents.

Besides the universal scaling function ${\tilde R(x,1)}$
the corresponding curve of an $\epsilon$-expansion obtained
from a renormalization group analysis is shown 
in Figs.\,\ref{fig:uni_scal_2d},\ref{fig:uni_scal_3d}.
Using the parametric representation~\cite{SCHOFIELD_1,JOSEPHSON_1}
of the absorbing
phase transition, Janssen {\it et al.} showed 
that the equation of state is given by the remarkably 
simple scaling function~\cite{JANSSEN_2} 
\begin{equation}
\label{eq:rg_equation_of_state}
H(\theta) \; = \; \theta \, (2-\theta) + {\cal O}(\epsilon^3),
\end{equation}
where $\epsilon$ denotes the distance to the
upper critical dimension $D_{\scriptscriptstyle \rm c}=4$, i.e., 
$\epsilon=D_{\scriptscriptstyle  \rm c}- D$.
Here the scaling behavior of the 
quantities $\rho_{\rm a}$, $\delta p$, and $h$ is
transformed to the variables $R$ and $\theta$
through the relations
\begin{equation}
\label{eq:para_transform}
b \, \delta p \; = \; R \, (1-\theta),
\quad\quad\quad
\rho_{\scriptscriptstyle \rm a} \; = \; R^\beta \, 
\frac{\theta}{2}. 
\end{equation}
The equation of state is given by 
\begin{equation}
\label{eq:para_transform_equa_state}
 a\, h\; = \; 
 \left ( \frac{R^\beta}{2} \right )^{\delta} 
 \; H(\theta) 
\end{equation}
with the metric factors $a$ and $b$.
The whole phase diagram is described by the parameter
range $R\ge 0$ and $\theta \in [ 0 , 2 ]$.
In Fig.\,\ref{fig:uni_scal_2d},\ref{fig:uni_scal_3d} 
a comparison between the numerically obtained scaling
functions and the analytical result of 
Eqs.\,(\ref{eq:rg_equation_of_state}-\ref{eq:para_transform_equa_state})
is made.
The RG-data differ slightly from the universal function.
As expected the differences decrease with increasing 
dimension and are especially strong in $D=1$~\cite{LUEB_27}.
This point is discussed in detail below.

\subsection{\textsf {Above the upper critical dimension}}

\begin{figure}[t]
  \centering
  \includegraphics[width=7cm,angle=0]{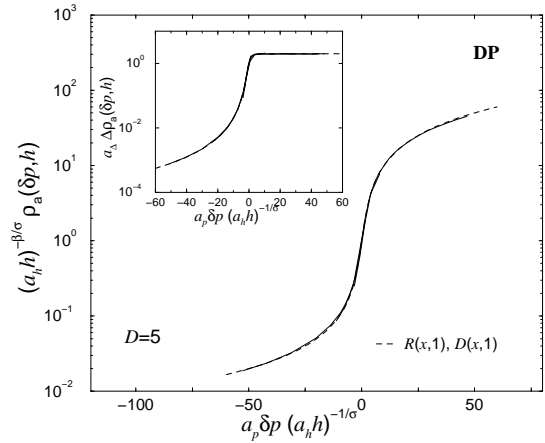}
  \caption{
    The universal scaling plots of the order parameter 
    and its fluctuations (inset) for $D=5$ and 
    for various values of the external field 
    ($h=5\,10^{-5}$, $7\,10^{-5}$, $10^{-6}$, $7\,10^{-7}$).
    The dashed lines correspond to the mean-field solutions
    ${\tilde R}_{\scriptscriptstyle {\mathrm {MF}}}(x,1)$ 
    and ${\tilde D}_{\scriptscriptstyle {\mathrm {MF}}}(x,1)$  
    [see Eqs.\,(\protect\ref{eq:mean_field_uni_order_scal},\protect\ref{eq:mean_field_uni_fluc_scal})].
   }
  \label{fig:uni_scal_5d} 
\end{figure}

Above the upper critical dimension the scaling
behavior of a phase transition equals the scaling behavior
of the corresponding mean-field solution 
[Eqs.\,(\ref{eq:uni_scal_mf_R}-\ref{eq:uni_scal_mf_C})].
Plotting  
$\rho_{\mathrm a}/\sqrt{a_{\scriptscriptstyle h} h}$ as a function
of ~$a_{\scriptscriptstyle p} \delta p/\sqrt{a_{\scriptscriptstyle h} h}$,
the numerical data should collapse to the universal 
scaling function
\begin{equation}
{\tilde R}_{\scriptscriptstyle {\mathrm {MF}}} 
(x, 1) \; = \;
\frac{x}{2} 
\, + \, \sqrt{1 \, + \, \left ( \frac{x}{2} \right )^2 \;}
\label{eq:mean_field_uni_order_scal}
\end{equation}
with the scaling variable 
$x=a_{\scriptscriptstyle p}\, \delta p \, / \, \sqrt{a_{\scriptscriptstyle h}\, h}$.
In Fig.\,\ref{fig:uni_scal_5d} we plot the 
corresponding rescaled data of the five-dimensional model.
A perfect collapse of the numerical data and ${\tilde R}(x, 1)$
is obtained. 
This is a confirmation of the RG-result
$D_{\scriptscriptstyle \mathrm c} =4$~\cite{OBUKHOV_2,CARDY_1}.
To the best of our knowledge no numerical evidence that  
five-dimensional DP exhibits mean-field scaling
behavior was published so far.

The rescaled fluctuation data is presented in
 Fig.\,\ref{fig:uni_scal_5d}.
As for the universal order parameter, 
the data of the fluctuations are in
agreement with the corresponding universal mean-field 
scaling function
\begin{equation}
{\tilde D}_{\scriptscriptstyle {\mathrm {MF}}} 
(x, 1) \; = \;  1 \, + \, 
\frac{x}
{\,2 \; \sqrt{1 \, + \, \left ( {x}/{2} \right )^2 \;}\, } .
\label{eq:mean_field_uni_fluc_scal}
\end{equation}


\subsection{\textsf {At the upper critical dimension}}

At the upper critical dimension $D_{\scriptscriptstyle \mathrm c}=4$
the scaling behavior is governed by the mean-field
exponents modified by logarithmic corrections.
For instance the order parameter obeys in 
leading order
\begin{eqnarray}
\label{eq:scal_OPzf_dc}
\rho_{\scriptscriptstyle \mathrm a}(\delta p, h=0) 
\; \propto \; 
\delta p \, | \ln{ \delta p}|^{\mathrm B}  \, , \\[2mm]
\label{eq:scal_OPcp_dc}
\rho_{\scriptscriptstyle \mathrm a} (\delta p=0, h) 
\; \propto \; 
\sqrt{ h}
\, | \ln{h}|^{\Sigma} \, .
\end{eqnarray}
The logarithmic correction exponents $\mathrm{B}$ and 
$\Sigma$ are characteristic features of the whole
universality class similar to the usual critical exponents.
Numerous theoretical, numerical, as well as experimental
investigations of critical systems at $D_{\scriptscriptstyle \mathrm c}$ 
have been performed (see for 
instance~\cite{LARKIN_1,AHARONY_1,GRIFFIN_1,BRINKMANN_1,AKTEKIN_1,LUEB_5,LUEB_10,LUEB_17,JANSSEN_3,JANSSEN_5,STENULL_1}).
Logarithmic corrections make the data analysis
quite difficult.
Hence most investigations are focused
on the determination of the correction exponents only, 
lacking the
determination of the scaling functions 
at $D_{\scriptscriptstyle \mathrm c}$.

Recently, a method of analysis was developed to 
determine the universal scaling functions at the
upper critical dimension~\cite{LUEB_26}.
In this work the authors use 
the phenomenological scaling ansatz (all terms in
leading order)
\begin{equation}
a_{\scriptscriptstyle \mathrm a}  \, \rho_{\scriptscriptstyle \mathrm a}(\delta p, h) 
\; \sim \; 
\lambda^{- \beta_{\scriptscriptstyle {\mathrm {MF}}}}\, | \ln{\lambda}|^{l} 
\; {\tilde R}
(a_{\scriptscriptstyle p}  
\delta p \; \lambda \, | \ln{\lambda}|^{b} , 
a_{\scriptscriptstyle h} h \;
\lambda^{\sigma_{\scriptscriptstyle {\mathrm {MF}}}}\, | \ln{\lambda}|^{s}) \, ,
\label{eq:uni_scal_EqoS_dc}
\end{equation}
with $\beta_{\scriptscriptstyle {\mathrm {MF}}}=1$ 
and $\sigma_{\scriptscriptstyle {\mathrm {MF}}}=2$.
Therefore, the order parameter at zero field ($h=0$) and at the
critical density ($\delta p=0$) are given in leading order by
\begin{eqnarray}
\label{eq:uni_scal_OPzf}
a_{\scriptscriptstyle \mathrm a}  \, \rho_{\scriptscriptstyle \mathrm a}(\delta p, h=0) 
& \sim & 
a_{\scriptscriptstyle p}  
\delta p \, | \ln{a_{\scriptscriptstyle p}  \delta p}|^{\mathrm B} 
\; {\tilde R}(1,0) , \\[2mm]
\label{eq:uni_scal_OPcp}
a_{\scriptscriptstyle \mathrm a}  \, \rho_{\scriptscriptstyle \mathrm a} (\delta p=0, h) 
& \sim & 
\sqrt{a_{\scriptscriptstyle h}  h}
\, | \ln{\sqrt{a_{\scriptscriptstyle h} h}}|^{\Sigma} 
\; {\tilde R}(0,1)
\end{eqnarray}
with ${\mathrm B}=b+l$ and $\Sigma=s/2+l$.
Similar to the case $D \neq D_{\mathrm c}$ 
the normalization ${\tilde R}(0,1)={\tilde R}(1,0)=1$
was used. 
According to the ansatz Eq.\,(\ref{eq:uni_scal_EqoS_dc})
the scaling behavior of the equation of state 
is given in leading order by
\begin{equation}
a_{\scriptscriptstyle \mathrm a}  \, \rho_{\scriptscriptstyle \mathrm a}(\delta p, h) 
\; \sim \; 
\sqrt{a_{\scriptscriptstyle h}  h}
\; | \ln{\sqrt{a_{\scriptscriptstyle h} h}}|^{\Sigma} 
\; {\tilde R}(x,1) 
\label{eq:uni_scal_EqoS}
\end{equation}
where the scaling argument is given by 
\begin{equation}
x \; = \; 
a_{\scriptscriptstyle p} \delta p \,
\sqrt{a_{\scriptscriptstyle h} h\,}^{\, -1} \,
| \ln{\sqrt{a_{\scriptscriptstyle h}  h}\,}|^{\Xi} 
\label{eq:uni_scal_arg_x}
\end{equation}
with $\Xi=b-s/2={\mathrm B}-\Sigma$.

In case of directed percolation it is possible to
confirm the scaling ansatz Eq.\,(\ref{eq:uni_scal_EqoS_dc}) 
by a RG-approach~\cite{JANSSEN_3}.
In particular the logarithmic correction exponents are 
given by $l=7/12$, $b=-1/4$, and $s=-1/2$.
Thus the scaling behavior of the equation of state
is determined by the logarithmic correction exponents~\cite{JANSSEN_3}
\begin{equation}
{\mathrm B} \; = \; \Sigma \; = \; 1/3 \, , 
\quad \quad  \quad 
\Xi=0 \, .
\label{eq:log_corr_exp_dp}
\end{equation}
It is worth mentioning that in contrast to the RG results
below the upper critical dimension the logarithmic
correction exponents do not rely on approximation
schemes like $\epsilon$- or $1/n$-expansions.
Within the RG theory they are exact results.

Similarly to the order parameter the following form
is used for its fluctuations~\cite{LUEB_26}
\begin{equation}
 a_{\scriptscriptstyle \Delta}  \; \Delta\rho_{\scriptscriptstyle \mathrm a}(\delta p, h) 
\; \sim \; 
\lambda^{\gamma^{\prime}}  \; | \ln{\lambda}|^{k} 
\; \; {\tilde D}
(a_{\scriptscriptstyle p}  
\delta p \; \lambda \, | \ln{\lambda}|^{b} , 
a_{\scriptscriptstyle h} h \;
\lambda^{-\sigma}\, | \ln{\lambda}|^{s}) \, .
\label{eq:uni_scal_fluc_dc}
\end{equation}
Using the mean-field value $\gamma^{\prime}=0$
and taking into account that numerical simulations
show that the fluctuations remain finite at the critical point
(i.e.~$k=0$) the scaling function
\begin{equation}
a_{\scriptscriptstyle \Delta}  \; \Delta\rho_{\scriptscriptstyle \mathrm a}(\delta p, h) 
\; \sim \; 
\; {\tilde D}(x,1)
\label{eq:uni_scal_Fluc}
\end{equation}
is obtained, where the scaling argument~$x$ is given 
by Eq.\,(\ref{eq:uni_scal_arg_x}) with $\Xi = 0$.
The non-universal metric factor $a_{\scriptscriptstyle \Delta}$
is determined again by the condition ${\tilde D}(0,1)=1$.

\begin{figure}
  \centering
  \includegraphics[width=7cm,angle=0]{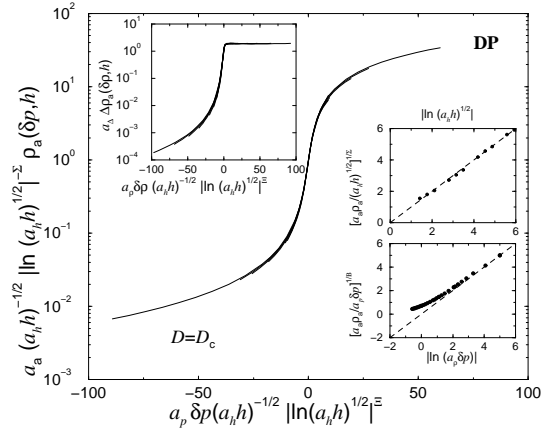}
  \caption{
    The universal scaling plots of the order parameter 
    and its fluctuations (upper left inset) at the upper critical dimension
    $D_{\scriptscriptstyle \mathrm c}=4$
    for various values of the external field 
    ($h=5\,10^{-5}$, $2\,10^{-5}$, $8\,10^{-6}$, $4\,10^{-6}$, $2\,10^{-6}$).
    The logarithmic correction exponents are given by
    ${\mathrm B} = \Sigma = 1/3$~\protect\cite{JANSSEN_3} and $\Xi=0$.
    The right insets show the order parameter at the critical density 
    and for zero field, respectively.
    The order parameter is rescaled according to 
    Eqs.\,(\protect\ref{eq:uni_scal_OPzf},\protect\ref{eq:uni_scal_OPcp}).
    Approaching the transition point 
    ($h\to 0$ and $\delta p \to 0$)
    the data tend to the function
    $f(x)=x$ (dashed lines) as required.
   }
  \label{fig:uni_scal_4d} 
\end{figure}

Thus the scaling behavior of the order parameter and 
its fluctuations at $D_{\scriptscriptstyle \mathrm c}$
is determined by two exponents (${\mathrm B}=1/3$
and $\Sigma=1/3$) and four unknown non-universal metric factors
($a_{\scriptscriptstyle \mathrm a},a_{\scriptscriptstyle p},
a_{\scriptscriptstyle h}, a_{\scriptscriptstyle \Delta}$).
Following~\cite{LUEB_26} we determine these values in our 
analysis by several conditions which are applied simultaneously:
first, both the rescaled equation of state and the rescaled
order parameter fluctuations have to collapse to the 
universal functions ${\tilde R}(x,1)$ and ${\tilde D}(x,1)$.
Second, the order parameter behavior at zero field and at the 
critical density are asymptotically given by the simple
function $f(x)=x$ when plotting 
$[a_{\scriptscriptstyle \mathrm a} \rho_{\scriptscriptstyle \mathrm a}(\delta p,0)
/a_{\scriptscriptstyle p} \delta p]^{1/{\mathrm B}}$
vs.~$|\ln{a_{\scriptscriptstyle p} \delta p}|$
and 
$[a_{\scriptscriptstyle \mathrm a} \rho_{\scriptscriptstyle \mathrm a}(0,h)
/\sqrt{a_{\scriptscriptstyle h} h\,}]^{1/{\Sigma}}$
vs.~$|\ln{\sqrt{a_{\scriptscriptstyle h} h\,}}|$,
respectively.
Applying this analysis we observe convincing results
for ${\mathrm B}=\Sigma=1/3$, $\Xi=0$, and  
for the values of the non-universal metric factors listed 
in Table\,\ref{table:metric_factors}.
The corresponding plots are presented in
Fig.\,\ref{fig:uni_scal_4d}.

\section{\textsf {UNIVERSAL AMPLITUDE COMBINATIONS}}
\label{sec:uni_ampl_comb}

In the following we consider several universal 
amplitude combinations
(see~\cite{PRIVMAN_2} for an excellent review).
As pointed out in~\cite{PRIVMAN_2},
these amplitude combinations are very useful in order
to identify the universality class of a phase transition
since the amplitude combinations vary more widely than
the corresponding critical exponents.
Furthermore, the measurement of amplitude combinations
in experiments or simulations yields a 
reliable test for theoretical predictions.
In particular, estimates of amplitude combinations
are provided by RG approximation schemes like $\epsilon$-
or $1/n$-expansions.

Usually numerical investigations focus
on amplitude combinations arising from finite-size
scaling analysis.
A well known example is the value of Binder's fourth
order cumulant at criticality~(see e.g.~\cite{BINDER_1}).
Instead of those finite-size properties we continue
to focus our attention to bulk critical behavior
since bulk amplitude combinations are of great experimental 
interest.
Furthermore, they can be compared to RG-results~\cite{JANSSEN_2}.

\begin{figure}[t]
  \centering
  \includegraphics[width=7cm,angle=0]{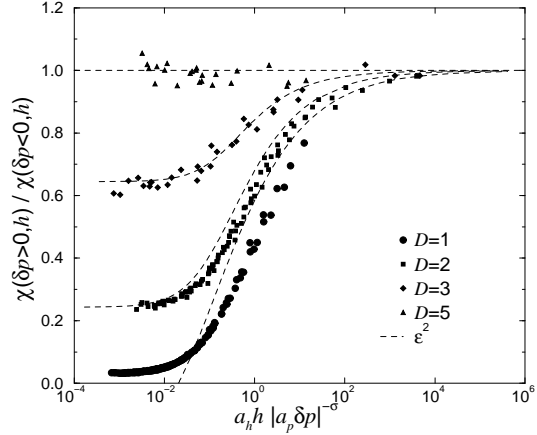}
  \caption{
    The universal scaling function ${\tilde C}(1,x)/{\tilde C}(-1,x)$ 
    for various dimensions.
    The dashed lines correspond to an $\epsilon$-expansion of a
    RG approach~\protect\cite{JANSSEN_2}.
    The universal amplitude ${\tilde C}(1,0)/{\tilde C}(-1,0)$ 
    is obtained from the extrapolation 
    $a_{\scriptscriptstyle h} h |a_{\scriptscriptstyle p} \, \delta p|^{-\sigma} \to 0$.
   }
  \label{fig:moment_sus} 
\end{figure}

The susceptibility diverges as
\begin{eqnarray}
\chi(\delta p >0, h=0) & \sim &
a_{\scriptscriptstyle \chi,+} \; \delta p^{-\gamma} \, , \\[2mm]
\chi(\delta p <0, h=0) & \sim &
a_{\scriptscriptstyle \chi,-} \; (-\delta p)^{-\gamma} \, ,
\end{eqnarray}
if the critical point is approached from above
and below, respectively.
The amplitude ratio
\begin{equation}
\frac{\chi(\delta p >0, h=0)}{\chi(\delta p <0, h=0)}
\; = \; \frac{\, a_{\scriptscriptstyle \chi,+}\, }{a_{\scriptscriptstyle \chi,-}}
\label{eq:class_ampl_sus}
\end{equation}
is a universal quantity similar to the critical exponents, 
i.e., all systems belonging to a given universality class 
are characterized by the same 
value $a_{\scriptscriptstyle \chi,+}/a_{\scriptscriptstyle \chi,-}$.
This can be seen from Eq.\,(\ref{eq:scal_ansatz_Susc}).
Setting $a_{\scriptscriptstyle p} | \delta p | \lambda=1$ yields 
\begin{equation}
\frac{\chi(\delta p >0, h)}{\chi(\delta p <0, h)}
\; = \; \frac{{\tilde C}(+1,x) }{{\tilde C}(-1,x)}
\label{eq:ampl_ratio_suscept}
\end{equation}
with $x=a_{\scriptscriptstyle h} h 
| a_{\scriptscriptstyle p} \delta p|^{-\sigma}$.
Obviously this is a universal quantity for all values
of the scaling variable $x$.
In particular it equals the 
ratio $a_{\scriptscriptstyle \chi,+}/a_{\scriptscriptstyle \chi,-}$
for $x \to 0$, i.e., vanishing external field.
In general, universal amplitude combinations are related to
particular values of the universal scaling functions.

In Fig.\,\ref{fig:moment_sus} the universal susceptibility
ratio Eq.\,(\ref{eq:ampl_ratio_suscept}) is shown 
for various dimensions.
The corresponding data saturates 
for $x\to 0$.
Our estimates for the amplitude 
ratios ${\tilde C}(+1,0)/{\tilde C}(-1,0)$ are
$0.033 \pm 0.004$ for $D=1$,
$0.25 \pm 0.01$ for $D=2$, as well as
$0.65 \pm 0.03$ for $D=3$.
In case of five-dimensional DP the  
amplitude ratio is constant, as
predicted from the mean-field behavior
\begin{equation}
\frac{{\, \tilde C}_{\scriptscriptstyle {\mathrm {MF}}}(+1,x) \, }
{{\tilde C}_{\scriptscriptstyle {\mathrm {MF}}}(-1,x)}
\; = \; 1 
\label{eq:ampl_ratio_suscept_mf}
\end{equation}
for all $x$.
The behavior of the ratio ${\tilde C}(+1,x)/{\tilde C}(-1,x)$ 
for $D< D_{\scriptscriptstyle \mathrm c}$ reflects the
crossover from mean-field to non mean-field 
behavior.
Far away from the transition point, the critical fluctuations
are suppressed and the behavior of the system is well 
described by the mean-field solution [Eq.\,(\ref{eq:ampl_ratio_suscept_mf})].
Approaching criticality the critical fluctuations
increase and a crossover to the $D$-dimensional behavior
takes place.

In the already mentioned work~\cite{JANSSEN_2}, 
Janssen {\it et al.}~calculated
the steady state scaling behavior of DP within a RG approach.
In particular they obtained for the susceptibility
amplitude ratio
\begin{equation}
\frac{{\tilde C}(+1,0)}{\,{\tilde C}(-1,0)\,}
\; = \;
1 \, - \, \frac{\epsilon}{\, 3 \, } \,
\left [
\, 1 \, - \,
\left (
\frac{11}{288} \, - \, \frac{53}{144} \, \ln{\frac{4}{3}} \,
\right ) \, \epsilon 
\, + \, {\cal O}(\epsilon^2)
\right ]
\label{eq:susc_ampl_ratio_epsilon}
\end{equation}
leading to $-0.2030\ldots$ for $D=1$,
$0.2430\ldots$ for $D=2$,
$0.6441\ldots$ for $D=3$.
Except for the unphysical one-dimensional result
these values agree well with our numerical estimates.

Furthermore the parametric representation of the 
susceptibility was derived in~\cite{JANSSEN_2}
and it is straightforward to calculate 
the universal ratio Eq.\,(\ref{eq:ampl_ratio_suscept}).
The results are plotted for various dimensions
in Fig.\,\ref{fig:moment_sus}.
It is instructive to compare these results
with the numerical data
since the theoretical curve reflects the accuracy
of the RG estimations of all three quantities,
the exponent, the scaling function, as well as the
non-universal metric factors.
All quantities are well approximated for the
three-dimensional model.
In the two-dimensional case we observe a horizontal
shift between the numerical data and the RG-estimates.
Thus the RG-approach yields good estimates for the exponents
and the scaling function but the metric factors are
of significantly less quality.
For $D=1$ the $\epsilon^2$-approximation does not 
provide appropriate estimates of the DP scaling
behavior.
Thus higher orders than ${\cal O}(\epsilon^2)$ are
necessary to describe the scaling behavior of directed percolation
in low dimensions.

\begin{figure}[b]
  \centering
  \includegraphics[width=7cm,angle=0]{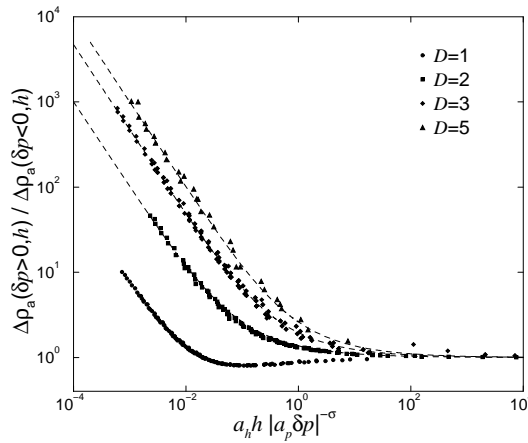}
  \caption{
    The universal scaling function 
    ${\tilde D}(1,x)/{\tilde D}(-1,x)$
    for various dimensions.
    The dashed line corresponds to the mean-field
    scaling behavior.
    For $D=2$ and $D=3$ the mean-field curves are shifted 
    the by the factors $9.56$ and $2.12$ to the left. 
   }
  \label{fig:moment_fluc} 
\end{figure}

Analogous to the susceptibility the 
universal amplitude ratio of the fluctuations
is given by
\begin{equation}
\frac{\Delta\rho_{\mathrm a}(\delta p >0, h)}
{\Delta\rho_{\mathrm a}(\delta p <0, h)}
\; = \; \frac{{\tilde D}(+1,x) }{{\tilde D}(-1,x)}
\label{eq:ampl_ratio_fluc}
\end{equation}
with $x=a_{\scriptscriptstyle h} h 
| a_{\scriptscriptstyle p} \delta p |^{-\sigma}$.
In the case of absorbing phase transitions this 
ratio diverges for vanishing field.
For $\delta p <0$ the order parameter
fluctuations are zero (absorbing state) for vanishing
field whereas the fluctuations remain finite above the
transition ($\delta p >0$).
Thus absorbing phase transitions are generally 
characterized by 
\begin{equation}
\frac{{\tilde D}(+1,0) }{{\tilde D}(-1,0)} 
\; \to \; \infty .
\label{eq:ampl_ratio_fluc_zero_field}
\end{equation}
In Fig.\,\ref{fig:moment_fluc} we plot the fluctuation
ratio [Eq.\,(\ref{eq:ampl_ratio_fluc})] as a function
of the scaling variable $a_{\scriptscriptstyle h} h 
| a_{\scriptscriptstyle p} \delta p |^{-\sigma}$
for various dimensions.
We observe in all cases that the fluctuation ratios 
diverge for $x \to 0$.
Only the one-dimensional system exhibits a particular
behavior characterized by the minimum of the 
corresponding curve.
The origin of this behavior is that for $D=2,3$ 
the universal scaling function ${\tilde D}(x,1)$ exhibits 
a maximum for $x>0$, whereas for $D=1$ 
it is located at $x<0$ (see Fig.\,8 in~\cite{LUEB_27}).

In the five-dimensional model we observe a perfect
agreement with the mean-field behavior
\begin{equation}
\frac{\, {\tilde D}_{\scriptscriptstyle {\mathrm {MF}}}(+1,x) \,}
{{\tilde D}_{\scriptscriptstyle {\mathrm {MF}}}(-1,x)}
\; = \; \frac{\,\phantom{-\,}1\,+\,\sqrt{1+4x\,}\,}
{\,-\,1\,+\,\sqrt{1+4x\,}\,} 
\; \; \mathop{\longrightarrow}\limits_{x\to 0} \; \; 
\frac{\,1+2x\, }{2 x} \, .
\label{eq:ampl_ratio_fluc_mf}
\end{equation}
Surprisingly, the two- and three-dimensional data are also
well approximated by this formula provided that 
one performs a simple rescaling ($x  \mapsto a_{\scriptscriptstyle D} x$)
which results in Fig.\,\ref{fig:moment_fluc}
in a horizontal shift of the data. 
We suppose that this behavior could be explained
by a RG-analysis of the fluctuations.

\begin{figure}[b]
  \centering
  \includegraphics[width=7cm,angle=0]{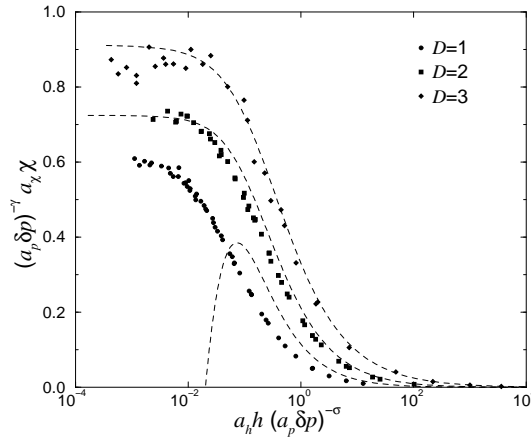}
  \caption{
    The universal scaling function ${\tilde C}(1,x)$ 
    for various dimensions.
    The dashed lines correspond to an $\epsilon$-expansion of an
    RG approach~\protect\cite{JANSSEN_2}.
    The universal amplitude $R_{\scriptscriptstyle \chi}$ 
    is obtained from the extrapolation 
    $a_{\scriptscriptstyle h} h (a_{\scriptscriptstyle p} \, \delta p)^{-\sigma} \to 0$.
   }
  \label{fig:C_1x} 
\end{figure}

Similar to the universal amplitude ratios 
of the susceptibility and the fluctuations
other universal combinations can be defined.
Well known from equilibrium phase transitions 
is the quantity (see e.g.~\cite{PRIVMAN_2}) 
\begin{equation}
R_{\chi} \; = \;
\Gamma \, d_{\scriptscriptstyle \mathrm c} \, B^{\delta-1} \, ,
\label{eq:R_chi}
\end{equation}
which is also experimentally accessible, e.g.~for
magnetic systems.
Here, $\Gamma$ denotes the amplitude of the 
susceptibility $\chi$ in zero field 
($\chi \sim \Gamma \, \delta T^{-\gamma}$) and
$B$ is the corresponding amplitude of the order
parameter $M$ ({$M \sim B \, \delta T^{\beta}$}).
The factor $d_{\scriptscriptstyle \mathrm c}$ describes
how the order parameter $M$ depends on 
the conjugated field $H$ at $\delta T=0$
($H \sim d_{\scriptscriptstyle \mathrm c} \, M^{\delta}$).

In case of directed percolation
these amplitudes correspond to the values
$B=a_{\scriptscriptstyle p}^{\beta}\, {\tilde R}(1,0)$,
$\Gamma = a_{\scriptscriptstyle p}^{\gamma} \,
a_{\scriptscriptstyle h} \, {\tilde C}(1,0) $
as well as
$d_{\scriptscriptstyle \mathrm c} = a_{\scriptscriptstyle h}^{-1}\, 
{\tilde R}(0,1)^{-\delta}$ where $\delta=\sigma/\beta$.
The normalizations ${\tilde R}(1,0)={\tilde R}(0,1)=1$ 
yield for the amplitude combination 
\begin{equation}
R_{\chi} \; = \;
{\tilde C}(1,0) 
\label{eq:R_chi_uni_equi}   
\end{equation}
which is obviously a universal quantity.
In Fig.\,\ref{fig:C_1x} 
the scaling function ${\tilde C}(1,x)$ is 
plotted as a function of 
$x=a_{\scriptscriptstyle h} h\, (a_{\scriptscriptstyle p} \, \delta p)^{-\sigma}$
for $D=1,2,3$.
The corresponding data saturates for $x\to 0$. 
Our estimates are 
$R_{\chi}=0.60\pm0.04$ for $D=1$,
$R_{\chi}=0.72\pm0.04$ for $D=2$, and
$R_{\chi}=0.86\pm0.08$ for $D=3$.
Note that the error-bars reflect only the data scattering
in  Fig.\,\ref{fig:C_1x}.
In contrast to the amplitude ${\tilde C}(1,0)/{\tilde C}(-1,0)$
the data of $R_{\chi}$ are affected by the uncertainties
of the exponent $\gamma$ and the uncertainties of the
metric factors $a_{\scriptscriptstyle p}$, $a_{\scriptscriptstyle h}$.
These uncertainties increase the error-bars significantly.
The two- and three-dimensional data
agree quite well with the RG-results
$R_{\chi}=0.7244\ldots\;$ for $D=2$ and
$R_{\chi}=0.9112\ldots\;$ for $D=3$~\cite{JANSSEN_2}.
In the one-dimensional model the $\epsilon^2$-expansion
yields again an unphysical result ($R_{\chi}=-3.927\ldots\;$).

\section{\textsf {CONCLUSIONS}}
\label{sec:conc}

We considered the universal steady state scaling 
behavior of directed percolation with an
external field in $D \ge 2$ dimensions. 
Our data for D=5 coincide with the mean field solution,
confirming that $D_{\scriptscriptstyle \mathrm c}=4$ is 
the upper critical dimension. 
At $D_{\scriptscriptstyle \mathrm c}$ we presented for the first 
time a numerical scaling analysis of DP including logarithmic 
corrections. 
Our results agree well with those of a recently performed
RG approach~\cite{JANSSEN_3}.
Apart from the scaling functions we also considered amplitude ratios 
and combinations for the order parameter fluctuations 
and the susceptibility. 
A comparison with RG~\cite{JANSSEN_2} results reveals that higher orders
than ${\cal O}(\epsilon^2)$ are necessary to describe
the scaling behavior in low dimensions.

We would like to thank H.-K.~Janssen and O.~Stenull for fruitful
discussions and for communicating their results
prior to publication.
Furthermore, we thank J.~Adler and S.~Kwon for helpful discussions
and useful comments on the manuscript.
R.\,D.~Willmann thanks the Weizmann Institute for warm hospitality
and the Einstein Center for financial support during a visit when 
this work was in progress.
S.~L\"ubeck thanks the Minerva Foundation (Max Planck Gesellschaft)
for financial support.


\begin{thebibliography}{10}

\bibitem{STANLEY_1}
{H.\,E.~Stanley}, Rev.~Mod.~Phys. {\bf 71},  S358  (1999).

\bibitem{GUGGENHEIM_1}
{E.\,A.~Guggenheim}, J.~Chem.~Phys. {\bf 13},  253  (1945).

\bibitem{MILOSEVIC_2}
S. Milo{\u {s}}evi{\'{c}} and {H.\,E.~Stanley}, Phys.~Rev.~B {\bf 6},  1002
  (1972).

\bibitem{PRIVMAN_2}
V. Privman, {P.\,C.~Hohenberg}, and A. Aharony,  in {\em Universal
  critical-point amplitude relations {\it in Phase Transitions and Critical
  Phenomena}, Vol.\,14}, edited by C. Domb and {J.\,L.~Lebowitz} (Academic
  Press, London, 1991).

\bibitem{WILSON_1}
{K.\,G.~Wilson}, Phys.~Rev.~B {\bf 4},  3174  (1971).

\bibitem{WILSON_2}
{K.\,G.~Wilson}, Phys.~Rev.~B {\bf 4},  3184  (1971).

\bibitem{WEGNER_1}
{F.\,J.~Wegner} and {E.\,K.~Riedel}, Phys.~Rev.~B {\bf 7},  248  (1973).

\bibitem{WILSON_3}
{K.\,G.~Wilson} and J. Kogut, Phys.~Rep. {\bf 12C},  75  (1974).

\bibitem{HINRICHSEN_1}
H. Hinrichsen, Adv.~Phys. {\bf 49},  815  (2000).

\bibitem{JANSSEN_1}
{H.\,K.~Janssen}, Z.~Phys.~B {\bf 42},  151  (1981).

\bibitem{GRASSBERGER_2}
P. Grassberger, Z.~Phys.~B {\bf 47},  365  (1982).

\bibitem{ROSSI_1}
M. Rossi, R. Pastor-Satorras, and A. Vespignani, Phys.\,Rev.\,Lett. {\bf 85},
  1803  (2000).

\bibitem{ESSAM_1}
{J.\,W.~Essam}, J.~Phys.~A {\bf 22},  4927  (1989).

\bibitem{CARDY_2}
{J.\,L.~Cardy} and {U.\,C.~T{\protect\"a}uber}, Phys.~Rev.~Lett. {\bf 13},
  4780  (1996).

\bibitem{JENSEN_2}
I. Jensen, Phys.~Rev.~Lett. {\bf 70},  1465  (1993).

\bibitem{JENSEN_3}
I. Jensen and R. Dickman, Phys.~Rev.~E {\bf 48},  1710  (1993).

\bibitem{LUEB_27}
S. L{\protect\"u}beck and {R.\,D.~Willmann}, J.~Phys.~A {\bf 35},  10205
  (2002).

\bibitem{MUNOZ_1}
{M.\,A.~Mu\~{n}oz}, G. Grinstein, R. Dickman, and R. Livi, Phys.~Rev.~Lett.
  {\bf 76},  451  (1996).

\bibitem{WIJLAND_1}
{F.~van\,Wijland}, Phys. Rev. Lett. {\bf 89},  190602  (2002).

\bibitem{GRASSBERGER_4}
P. Grassberger and {A.~de\,la\,Torre}, Ann.~Phys.~(N.Y.) {\bf 122},  373
  (1979).

\bibitem{PFEUTY_1}
P. Pfeuty and G. Toulouse, {\em Introduction to the renormalization group and
  critical phenomena} (John Wiley \protect\,\&\,Sons, Chichester, 1994).

\bibitem{JANSSEN_2}
{H.\,K.~Janssen}, {\protect\"U}. Kutbay, and K. Oerding, J.~Phys.~A {\bf 32},
  1809  (1999).

\bibitem{JANSSEN_3}
{H.\,K.~Janssen} and O. Stenull, Phys.~Rev.~E {\bf 69}, 016125
  (2004).

\bibitem{GRASSBERGER_6}
P. Grassberger, Phys.~Rev.~E {\bf 56},  3682  (1997).

\bibitem{GRASSBERGER_7}
P. Grassberger, R. Hegger, and L. Sch{\protect\"a}fer, J.~Phys.~A {\bf 27},
  7265  (1994).

\bibitem{LUEB_5}
S. L{\protect\"u}beck, Phys.~Rev.~E {\bf 58},  2957  (1998).

\bibitem{LUEB_10}
{D.\,V.~Ktitarev}, S. L{\protect\"u}beck, P. Grassberger, and
  {V.\,B.~Priezzhev}, Phys.~Rev.~E {\bf 61},  81  (2000).

\bibitem{LUEB_17}
L. Roters, S. L{\protect\"u}beck, and {K.\,D.~Usadel}, Phys.~Rev.~E {\bf 66},
  069901  (2002).

\bibitem{GRASSBERGER_5}
P. Grassberger, Phys.~Rev.~E {\bf 67},  036101  (2003).

\bibitem{LUEB_26}
S. L{\protect\"u}beck and {P.\,C.~Heger}, Phys.~Rev.~Lett. {\bf 90},  230601
  (2003).

\bibitem{FEDORENKO_1}
{A.\,A.~Fedorenko} and S. Stepanow, Phys.~Rev.~E {\bf 67},  057104  (2003).

\bibitem{JANSSEN_5}
{H.\,K.~Janssen} and O. Stenull, Phys.~Rev.~E {\bf 68},  036131  (2003).

\bibitem{STENULL_1}
O. Stenull and {H.\,K.~Janssen}, Phys.~Rev.~E {\bf 68},  036129  (2003).

\bibitem{DOMANY_1}
E. Domany and W. Kinzel, Phys.~Rev.~Lett. {\bf 53},  311  (1984).

\bibitem{TRETYAKOV_1}
{A.\,Y.~Tretyakov} and {N.~Inui}, J.~Phys.~A {\bf 28},  3985  (1995).

\bibitem{GRASSBERGER_3}
P. Grassberger and {Y.-C.~Zhang}, Physica~A {\bf 224},  169  (1996).

\bibitem{PRIVMAN_3}
V. Privman and {M.\,E.~Fisher}, Phys.~Rev.~B {\bf 30},  322  (1984).

\bibitem{MORI_1}
H. Mori and {K.\,J.~Mc\,Neil}, Prog.~Theor.~Phys. {\bf 57},  770  (1977).

\bibitem{JENSEN_5}
I. Jensen, J.~Phys.~A {\bf 32},  5233  (1999).

\bibitem{VOIGT_1}
{C.\,A.~Voigt} and {R.\,M.~Ziff}, Phys.~Rev.~E {\bf 56},  R6241  (1997).

\bibitem{JENSEN_6}
I. Jensen, Phys.~Rev.~A {\bf 45},  R563  (1992).

\bibitem{SCHOFIELD_1}
P. Schofield, Phys.~Rev.~Lett. {\bf 22},  606  (1969).

\bibitem{JOSEPHSON_1}
{B.\,D.~Josephson}, J.~Phys.~C {\bf 2},  1113  (1969).

\bibitem{OBUKHOV_2}
{S.\,P.~Obukhov}, Physica~A {\bf 101},  145  (1980).

\bibitem{CARDY_1}
{J.\,L.~Cardy} and {R.\,L.~Sugar}, J.~Phys.~A {\bf 13},  L423  (1980).

\bibitem{LARKIN_1}
{A.\,I.~Larkin} and {D.\,E.~Khmel'nitski{\u \i}}, JETP {\bf 29},  1123  (1969).

\bibitem{AHARONY_1}
A. Aharony, Phys.~Rev.~B {\bf 8},  3363  (1973).

\bibitem{GRIFFIN_1}
{J.\,A.~Griffiths}, {J.\,D.~Litster}, and A. Linz, Phys.~Rev.~Lett. {\bf 38},
  251  (1977).

\bibitem{BRINKMANN_1}
J. Bringmann, R. Courths, and {H.\,J.~Guggenheim}, Phys.~Rev.~Lett. {\bf 40},
  1286  (1978).

\bibitem{AKTEKIN_1}
N. Aktekin, J.~Stat.~Phys. {\bf 104},  1397  (2001).

\bibitem{BINDER_1}
K. Binder and {D.\,W.~Heermann}, {\em Monte Carlo Simulation in Statistical
  Physics} (Springer, Berlin, 1997).

\end{thebibliography}
\end{document}